\documentstyle[eqsecnum,aps]{revtex}

\begin{document}
\draft
\title{On the Topological Nature of Fundamental Interactions}
\author{M. Spaans\footnote{Hubble Fellow}}
\address{Harvard-Smithsonian Center for Astrophysics\\
60 Garden Street, Cambridge MA 02138, USA}
\date{\today}
\maketitle
\begin{abstract}
A thought experiment is proposed to unify quantum mechanics and general
relativity. The central paradigm is that space-time {\it topology} is
ultimately responsible for the Heisenberg uncertaintly principle and the
equivalence of inertial and gravitational mass.
It is found that Plankian space-time exhibits a complicated, but also
definite, multiply connected character. In this framework, an analysis of the
interactions in Nature is presented.

I.\ The Universal ground state of the constructed theory derives from the
properties of the topological
manifold $Q=2T^3\oplus 3S^1\times S^2$, which has 23 intrinsic degrees of
freedom, discrete $Z_3$ and $Z_2\times Z_3$ internal groups, an $SU(5)$
gauge group, and leads to a $U(1)$ symmetry on a lattice.
The structure of $Q$ provides a unique equation motion for the
mass-energy and particle rest mass wave functions. In its excited state
the Universe is characterized by a lattice of three-tori, $L(T^3)$.
The topological identifications present in this structure, a direct reflection
of the Heisenberg uncertainty principle, provide the boundary conditions for
solutions to the equation of motion, and suggest an interpretation for the
conceptually difficult concept of quantum mechanical entanglement.

II.\ In the second half of the paper the (observable) properties of $Q$ and
$L(T^3)$ are investigated. One reproduces the standard model,
and the theory naturally contains a Higgs field with possible
inflation. The thermodynamic properties of $Q$ yield a consistent amplitude
for the cosmic microwave background fluctuations, and the manifold $Q$
possesses internal energy scales which are independent of the field theory
defined on it, but which fix the predicted mass hierarchy of such
theories. The electron and its neutrino are identified as
particle ground states and their masses, together with those of all other known
particles, are predicted. A mass of $m_{\rm H}=131.6$ GeV is found for the
Higgs boson.

Furthermore, observational diagnostics are constructed which reflect the
underlying topology of Planckian space-time, and which are directly related to
phenomena on much larger scales. Specific predictions are made for the coupling
constants, quark confinement, black hole states, and the cosmological constant.
The latter is found to be almost zero. A heuristic argument for the occurrence
and magnitude of CP violation is given.

III.\ Finally, a discrete spectral signature is predicted at integer and
inverse integer multiples of the zero point frequency $\nu_0=857.3588$ MHz
(34.96698 cm). That is, each photon of frequency $\nu_0m$, for an integer $m$,
is paired with an otherwise identical photon $\nu_0/m$, produced by the vacuum,
but not vice versa. The discrete features $\nu_0m$ (and $\nu_0/m$) always have
a width of $7^{-3}\nu_0$ (and $7^{-3}\nu_0/m^2$), and should provide a direct
test of the reality of $Q$ and $L(T^3)$ independent of any standard model
physics.
\end{abstract}

\section{Introduction}

One of the outstanding questions in quantum cosmology and particle physics
is the unification of gravity with the electroweak and strong interactions.
Much effort has been devoted in the past years to formulate a purely
geometrical and topological theory for these types of interactions\cite{1,2}.
Probably best known are the theories involving 11-dimensional super-gravity and
superstrings\cite{1}. These theories have been shown recently to be
unified in
M-theory, although the precise formulation of the latter is not known yet.
Three outstanding problems in these approaches are compactification down to
four dimensions, the existence of a unique vacuum state, and most importantly
the formulation of a guiding Physical Principle to lead the mathematical
construction of the theory.

This work aims at using the mathematical properties of so-called prime nuclear
three-manifolds investigated in\cite{3}. These manifolds,
e.g.\ the three-torus $T^3$ and the handle $S^1\times S^2$, have the nice
property that they cannot be decomposed in other three-manifolds up to
homeomorphisms, and, to be called nuclear, that they
bound a Lorentz four-manifold with $SL(2;C)$ spin structure.
The three-sphere $S^3$ is also a prime manifold, but is not nuclear. The only
mathematical requirement to use these three-manifolds is that space-time can
be described by a (3+1)-dimensional topological manifold\cite{3}.

The present paper provides arguments which assign a fundamental meaning to the
three prime manifolds mentioned above. The driving philosophy is that
{\it topology} plays an essential role in the fundamental structure which
underlies Nature. Furthermore, since observations are the ultimate arbiter of
the correctness of a theory, specific predictions are made here which are
testable and which cannot be altered through any free parameters in the theory.

This paper is organized as follows. Section 2 presents a thought experiment
which relates quantum mechanics to the topology of space-time.
Section 3 discusses the general properties of a theory based on topological
manifolds constructed from $T^3$,
$S^1\times S^2$ and $S^3$, and Section 4
presents the derivation of the fundamental
equation of motion for the mass-energy and rest mass wave functions.
The equation of motion constitutes the main theoretical result of this work.
Section 5 discusses the particle
physics properties of the resulting theory and presents definite predictions
for the masses of particles. Section 6 investigates the cosmological
ramifications of solutions to the equation of motion. Section 7 presents an
analysis of the microstates of black holes and the cosmological constant.
Section 8 discusses the 
relationship between the constructed topological manifolds and the nature of
QCD as well as the values of the coupling constants.
Section 9 analyzes the nature of CP violation.
Section 10 presents a remarkable observational characteristic of
the theory in that there are discrete photon frequencies which are
accompanied by dual partners generated by the vacuum. Section 10 further
completes the results on the cosmological constant.
Section 11 contains the conclusions and discussion.

\section{Space-time Topology and the Nature of Forces}

In order to construct a physical theory of all known interactions,
one needs to formulate a Physical Principle
that accommodates both general relativity and quantum mechanics.
The theory resulting from the sections below aims at doing this, and will be
referred to as Topological Dynamics (TD) from hereon.

\subsection{Thought Experiment}

1. Imagine an observer with a measuring rod of accuracy $\ell$.
This same observer is located in a zero gravity
environment to witness the motion of an object which starts out at some initial
position. As he performs the same
experiment for smaller and smaller scales he notices that a smooth
description of the observational data is limited by the
accuracy of his measuring rod. The observed trajectories of various
objects still appear to be continuous but derivatives are very uncertain.
The most natural conclusion is that the paths are close to
identical within the error bars set by his measuring rod. Nevertheless,
any set of measurements, in principle, also leaves room
for the conclusion on his part that something prevents the particle
trajectories from becoming identical on scales $\le\ell$.

In the latter case, the physical obstruction responsible for the different
trajectories is hidden by the observer's inability to measure accurately. As
the observer repeats his experiments for all orientations in three-dimensional
space, he finds the same result. Therefore, his finite accuracy is formally
consistent with particle trajectories which are separated from one another by
the action of what appear to be loops (enclosed regions) when he projects his
data onto a hyperplane. These loops are a measure of the possible differences
between the particle histories, and have a size of the order of $\ell$.
Therefore, even though his measured
particle trajectories may very well be topologically trivial, the
observer has no definite way of confirming this possibility.

2. Imagine the observer in a satellite orbiting a black hole. To determine
the properties of the black hole and to see if his conclusions about particle
trajectories differ from the zero gravity case, he first establishes the
crucial concepts of space-time curvature and general covariance\footnote{The
occurrence of the force of gravity is a prediction of the work presented
here and is discussed in the Appendix.}.
The observer then measures the motion of probes in closer and closer orbits
and establishes the existence of an event horizon. When he performs
measurements within $\ell$ of the Schwarzschild radius, he again concludes
that there is an uncertainty as to the probe's actual trajectory. Formally,
his measurement accuracy does not allow him to exclude the possibility that his
probe moved into the horizon and back out, for scales of the order of $\ell$.
Even though the observer is disturbed by this fact, he can quite plausibly
hope that higher accuracy measurements will settle the issue. The observer can
confidentally conclude that there is a closed surface, and
only one generic loop of size $\sim\ell$ associated with particle trajectories
due to the focusing effect of space-time curvature close to the horizon.

3. Now {\it postulate} that the finite accuracy of the observer's measuring rod
is a property of Nature itself when $\ell =\ell_{\rm Planck}$, i.e.\ Nature
reaches these same conclusions {\it everywhere}.
The measurements then reflect an {\it intrinsic} property rather than an
observational external one, and the {\it possible} existence of topologically
distinct trajectories becomes a {\it requirement}. That is,
Nature quantizes particle trajectories on a scale
$\ell_{\rm Planck}$ and {\it distinguishes} these {\it paths} in a way that
requires only the continuity of the underlying space-time manifold.

\subsection{An Extension of Mach's Principle}

4. If one uses a mathematical framework
in which Planck scale space-time possesses the property of continuity, and is
locally flat when gravitational effects can be ignored, then one arrives at a
description where quantum mechanics and general relativity require the
combined presence of the flat three-torus and the curved handle. Both these
manifolds are nuclear and hence introduce a notion of length, chosen to be
$\ell_{\rm Planck}$, through the metric induced on the four-space they bound.
Because the handle contains a
closed $S^2$ surface, it provides a notion of $inside$ and $outside$, and hence
it provides the conjugate functions $\xi_{in}$ and $\xi_{out}$ that together
express the invariance under homeomorphisms $\xi_{in}\circ\xi_{out}=1$. Thus,
one finds that $\xi_{out}=\xi^{-1}_{in}$ and the pair (length,length$^{-1}$)
for the gravitational ($S^2$ is intrinsically curved) mass.
The non-trivial topological properties of these manifolds facilitate the
possible distinct particle trajectories discussed above.

5. The putative underlying topology should also be {\it global} in the
sense that the position of a particle is localized,
but is not exactly determined when one considers the Heisenberg uncertainty
principle. That is, Nature has to provide for all possible locations of a
particle, not just its expectation value. The way, based on the premise of
topologically distinct paths, in which to allow a {\it non-zero} probability
for a particle {\it anywhere} is to construct, for every time, a lattice of
three-tori, $L(T^3)$, through
$S^3$ surgery in which each three-torus is identified with any other. This
structure is automatically Lorentz invariant since $T^3$ is nuclear, and
provides four homotopically distinct paths between any two space-time
points\cite{3}. As such, one immediately obtains the superposition principle
in that a wave amplitude $\psi$ travelling along one of these four paths is
subject to all possible homeomorphisms. Furthermore, one can pair this path
$p_1$ with a second path $p_2$ to yield a notion of inertial mass for the
loop $S^1=p_1\circ p_2^{-1}$, like the argument for $S^2$ above.
Applying the same arguments to the remaining two paths, yields the
probability squared expression $\psi\psi^*$.

Furthermore, handles can naturally live on $L(T^3)$\cite{3}, and provide the
expression of the equivalence principle. That is, along any trajectory an
observer can ask the question: What is the magnitude of the inertial and
gravitational mass? Since Nature has to allow simultaneously for paths
that do and do not pass through handles, it cannot make any (quantitative)
distinction between inertial and gravitational mass, chosen to be
$m_{\rm Planck}$.

6. The underlying philosophy of TD is now that the information required to
represent {\it any} physical phenomenon, e.g.\ through paths, locations and
interactions, in Nature is stored in the {\it geometrical
and topological} degrees of freedom of space-time through $L(T^3)$ and the
handles, and that this specific space-time structure is fundamental, i.e.\ it
is the cause rather than the effect. As such, the argument is that
the intrinsic likelihood of any set of phenomena does not matter for the
underlying space-time structure which must accommodate them. Indeed, Nature has
to provide for all phenomena, not just the ones which are likely. Note in this
respect that the different trajectories contributing to the Feynman path
integral reflect, in this approach, the topological freedom in space-time
itself, and that the paths are considered fundamental. The above philosophy is
in line with the experimental results presented in \cite{4}.

7. One can now provide an extended physical basis for Mach's principle. The
question is how a particle knows which way to move under the influence of
inertial forces, if one rejects the notion of absolute space-time.
The thought experiment suggests that it is the combination of
{\it all possible} particle motions. As such, the geometry of space-time as
well as its topology, through the identifications in $L(T^3)$, determine how
particles move and interact. Because the thought experiment applies to
{\it any} time-like slice, it is the global 3+1 topology (just like it is the
global 3+1 geometry for Einstein gravity) which enters Mach's principle.

This notion of an underlying topology which reflects the Heisenberg uncertainty
principle provides a natural place for entanglement. The identifications on
$L(T^3)$ facilitate closed loops at any base point in space-time. If these
associated loops are linked, then Nature has at its disposal a means to
perceive, on a global scale, any two space-time points as distinct or related.
In fact, one could argue on this basis that the
maximum combined spatial and temporal dimension of the Universe is four
{\it if} one views entanglement as a requirement on Nature
to distinguish between linked and unlinked configurations. Linkage is a
topologically non-trivial invariant in three dimensions only.

The remainder of this paper is devoted to the way in which $L(T^3)$ and the
handles provide for fundamental interactions and physical parameters. One
should continuously bear in mind that the approach followed here is a
topological/geometrical one in the strict sense that space-time itself provides
both the stage for and the performers of physical phenomena.

\subsection{Mathematical Formulation}

After the above arguments, one can use the algebraic formulation
of\cite{3} to
quantify the number of degrees of freedom of the individual prime manifolds
and the structures defined by them. That is, the dynamics of the mathematical
theory are determined by the loop creation, $T^\dagger$, and loop annihilation,
$T$, operators, which obey $[T,T^\dagger ]=1$.
Their actions on a manifold $M$ are $T^\dagger M=S^1\times M$ and $TM=nM'$,
with $n$ the number of $S^1$ loops in $M$, and $M'$ the manifold $M$ with
a loop shrunk to a point.
The number of degrees of freedom of the prime manifolds, referred to as the
prime quanta, under the action of the scalar operator
$O=TT^\dagger +T^\dagger T\equiv A+B$, are 1, 3, and 7, for the three-sphere,
the handle manifold, and the three-torus, respectively.

The prime quanta have natural interpretations. The prime quantum of $T^3$
under $A+B=2A+1$ reflects the fact that in a lattice, the embedding in
four-space, with its intrinsic uncertainty $t_{\rm Planck}$, groups seven
three-tori (the one and its six neighbours) into one equivalence class.
In the case of $S^1\times S^2$, the non-zero Planck time causes the embedding
to group triplets of handles along the time axis because of the event horizon.

Since the three-tori only possess a scale, their prime quantum represents an
effective dimension of seven, i.e.\ every heptaplet of three-tori should be
viewed as an independent 7+3-dimensional entity. For the massive handles one
concludes that each facilitates, irrespective of position and size, the
{\it possible} spontaneous creation of a pair of mini black
holes every Planck time. Of course, this is a purely topological argument, and
it depends on the properties of the black hole system and the vacuum, whether
this possibility is realized.

Finally, in\cite{3} it
is shown that $L(T^3)$ supports an $SU(N)$ symmetry group,
where $2N=7+3$, because $L(T^3)$ naturally yields a self-interaction potential
$V=\mu^2\Phi^2+\lambda\Phi^4$, for constants $\mu$ and $\lambda$, and a
scalar\footnote{The vacuum expectation value of $\Phi$ must be Lorentz
invariant because of the three-torus.} multiplet $\Phi$. The maximum degree of
four in the interaction potential follows from the fact that there are four
homotopically inequivalent paths on $L(T^3)$ between any two points. These
paths meet {\it locally} in double and quadruple vertices. That is, the
potential $V$ is even since the field $\Phi$ is defined on $S^1$ loops.

The aim is now to construct a topological manifold
$Q$ which unifies the fundamental fields in Nature, and provides the
appropriate equation of motion, symmetry groups, coupling constants and
particle masses.

\section{The Ground State Manifold $Q$}

A manifold $Q=aT^3\oplus bS^1\times S^2$ which is built from three-tori and
handles, and has an odd number of constituents, is nuclear because odd sums of
nuclear primes bound Lorentz manifolds\cite{3}. The unification of fields
should be implemented topologically, if one accepts the central paradigm.
In this, the three-tori and handles represent quantum
field theory and gravity, respectively. A unification of the two should
therefore not distinguish one from the other. Hence, the number of paths
through both manifolds, representing the degrees of freedom, should be equal.
Note that we are unifying principles, not just fields. The paths through the
three-tori express the superposition principle in the presence of inertial
forces. The paths through the handles then facilitate the
equivalence principle, where mass-energy and space-time form a joint system.
The superposition principle always requires four degrees of freedom, prompting
a gravity theory with four degrees of freedom.

The number of distinct paths on $L(T^3)$, four between any two points, should
then be equal to the number of them provided by the $b$ handles, i.e.\
$a=2$ because any three-torus is identified with any other three-torus, and
$b+1=4$. That is, the unification proceeds through the enumeration of degrees
of freedom, in a spirit which views paths, like in the Feynman path integral,
as truly fundamental. Therefore, one finds the connected sum of three handle
manifolds and two three-tori, and hence $Q$ is Lorentz invariant,
$$Q=2T^3\oplus 3S^1\times S^2.\eqno(1)$$
This construction with three-tori and handles may make a handwaving impression
on the reader. In the Appendix it is shown that the result for $Q$ can be
derived in a self-consistent manner from first principles.

If one ignores the underlying $L(T^3)$ structure, then the presence of mini
black holes on the Planck scale is certainly not a new idea. Wheeler's approach
to geometrodynamics, and many other theoretical investigations, has emphasized 
this. The new element is the presence of the lattice of three-tori, which is
topologically very non-trivial, and introduces additional space-time
characteristics as well as a specific implementation of the mini black hole
idea.

The number of degrees of freedom $N_Q$ of $Q$ under the action of $O$ is
$$OQ=23Q.\eqno(2)$$
In the approach adopted here, these degrees of freedom are all
distinct and they reflect the different possible topological realizations of
$Q$. Because $Q$ is the topological expression of an arbitrary mass, an
arbitrary scale, {\it and} a gauge group, the degrees of freedom of $Q$ are
identified with different particle states.

\subsection{Latent Heat}

For the submanifold $P=T^3\oplus T^3$, one has $N_P=14$. The ``latent heat''
$H$ associated with the evaporation of the handle triplet,
$\Theta =3S^1\times S^2$, is therefore
$$H=(N_Q-N_P)m_{\rm Planck}/N_Q=9m_{\rm Planck}/23.\eqno(3)$$
Since the two three-tori in the structure $Q$ are identical objects, the
specific heat per three-torus is given by $h=H/2$.

\subsection{Specific Volume}

The $N_Q=23$ realizations of $Q$ and the $N_{T^3}=7$ of $T^3$ imply that the
``unit of length'' is $1/N_Q$ and $1/N_{T^3}$ on these topological manifolds,
respectively. The magnitude of the specific volumes in three-space of $Q$ and
$T^3$ are thus given by
$$\delta\rho /\rho (Q)=N_Q^{-3}=8.2\times 10^{-5},\eqno(4a)$$
$$\delta\rho /\rho (T^3)=N_{T^3}^{-3}=2.9\times 10^{-3}.\eqno(4b)$$

\subsection{General Properties of $Q$ and $L(T^3)$}

\subsubsection{Discrete Groups Generated by $P$ and $\Theta$}

The effective action $s^3$ of the handle triplet on $Q$ obeys
$$s^3=1.\eqno(5)$$
That is, a round trip along the manifold $\Theta$ necessarily picks up three
phases, which should add up to $2\pi$ since the loop algebra satisfies
$[T,T^\dagger ]=1$. Because all 3
handles are identical, this implies a $Z_3$ invariance for the individual
quantum fields in the theory defined on $Q$, with angles
$\theta_i=\{0,\pm 2\pi /3\}$. From the same arguments it follows that the
submanifold $P$ generates a $Z_2\times Z_3$ symmetry because one cannot
distinguish either three-torus in $P$.

It should be pointed out that precisely these symmetry groups have been
used in\cite{5} to
resolve the problem of doublet-triplet splitting, which results
from the unavoidable mixing of Higgs doublets $H,\bar{H}$ with their colored
triplet partners $T,\bar{T}$ and leads to an unacceptably rapid proton decay.
The above discrete groups in fact provide a natural resolution of the $\mu$
problem in terms of the topology of space-time\cite{5}.

\subsubsection{$U(1)$ Symmetries and Curvature}

The important distinction between $T^3$ and $L(T^3)$ is the presence of
junctions which connect the individual three-tori through $S^3$ surgery and
yield a lattice. The existence of junctions between the three-tori generates
a $U(1)$ symmetry, i.e.\ one can perform a twist along any junction without
changing the underlying topology.

Although $L(T^3)$ is above all a topologically complicated manifold, it does
provide a natural arena for general relativity\cite{3}. On scales much larger
than $\ell_{\rm Planck}$, or rather at sufficiently low energies, bending of
the $S^3$ lattice junctions can occur in accordance with Einstein gravity. The
ground state $Q$ should be viewed as encompassing general
relativity without making a decomposition in a curvature tensor and an
energy-momentum tensor (see Section 4.1).

\subsection{Particle Sectors on $Q$}

The homotopic properties of $Q$ should
lead to specific particle sectors. In this, the photon is not
viewed as being generated through the homotopic structure of $Q$, but results
from the junction degrees of freedom, i.e.\ the $U(1)$ twists. Note then
that the electromagnetic and gravitational field both have four degrees of
freedom, which are provided by the four homotopically distinct paths on
$L(T^3)$. Also, the Higgs boson is associated with the excitation of $Q$ to
$L(T^3)$.

The number of degrees of freedom $N_Q$ is the eigenvalue of the operator
$O=TT^\dagger +T^\dagger T\equiv A+B$ acting on $Q$.
There is then a natural division of the 23 degrees of freedom under
$AQ=14Q$ and $BQ=9Q$. Furthermore, the decomposition $OQ=O(P\oplus\Theta )$
has the same distribution of degrees of freedom under $A$ and $B$, and leads to
the further divisions
$$AQ=OP=(8+6)P,\eqno(6)$$
with eight plus six particles and
$$BQ=O\Theta =(3+6)\Theta ,\eqno(7)$$
with three plus six particles.
Clearly, the number of elementary particles is the same in both sectors, each
further partitioned in two triplets under $P\rightarrow T^3$. The number of
field particles then follows from the decomposition. Since $\Theta$ is broken,
it must possess a fully mixed neutral and a charged triplet under the action of
the $U(1)$ on the $L(T^3)$ junctions.

The junction potential on $L(T^3)$ or $Q$ supports the symmetry group
$SU(5)$. The $P$ and $\Theta$ sectors decompose $Q$ and are therefore
associated with subgroups. These subgroups can only contain $SU(N<5)$ and
$U(1)$ because of the form of the junction potential $V$ and the $U(1)$ twist
groups. For $SU(5)\sim SU(3)\times [SU(2)\times U(1)]$ these
constraints are satisfied, because $SU(3)$ contains 8 ($P$) field particles,
$SU(2)$ only 3 ($\Theta$), and there is one $U(1)$ junction on $Q$. Below it
will be shown that the field particles
must be bosonic. Therefore, $Q$ defines a ground state which
corresponds to the standard model.

\subsection{The Pauli Exclusion Principle and Topological Identifications}

A priori, both fermionic and bosonic sectors exist for the particle sectors
identified above, and one should address the question of
supersymmetry. That is, why does the Universe (not) distinguish between
fermions and bosons. When $\Theta$ evaporates, the structure $L(T^3)$
becomes the fundamental Lorentz invariant Planck scale object. Subsequently,
interactions are mediated by field particles which travel along the 6 junctions
surrounding any $T^3$. That is, it is the discrete three-torus with its
seven degrees of freedom under the operator $O$ which supports a field and
its quanta. Any field dynamics on $L(T^3)$ therefore requires the interaction
of two identical field particles on a three-torus. If these field particles are
manifestly fermionic, this violates the Pauli exclusion principle. Thus, only
bosonic field particles can carry the strong and electroweak force {\it and}
satisfy the Pauli exclusion principle on $L(T^3)$.

The origin of the Pauli exclusion principle actually {\it follows} from the
homotopic structure of $T^3$. It is easy to see\cite{2} that a spin $1/2$
particle requires two $S^1$ loops on a three-torus for its support. For two
identical fermions one thus finds that two $S^1$ loops are collapsed to one.
The consequence is that the three-torus becomes indistinguishable from the
prime manifold $S^1\times R_1$, with $R_1$ a Riemann surface of genus one. This
manifold is not nuclear and breaks Lorentz invariance. Therefore, one finds a
{\it topological} constraint which precludes interactions mediated by fermionic
field particles. Indeed, the discrete $Z_3$ and $Z_2\times Z_3$ symmetry
groups, discussed in \S 3.3.1 as a possible resolution of proton decay, simply
reflect this underlying topological obstruction.

Finally, the masses of any
inert, except for gravity, dual (in the sense of Section 10) partners are given
by $\surd (m_{\rm cel}m_{\rm H})$, with $m_{\rm H}$ and $m_{\rm cel}$ the
masses of the Higgs boson and the {\it charged} elementary particles,
respectively. Non-zero charge is required because the $U(1)$ on the $L(T^3)$
lattice junctions cannot prevent the violation of the topological particle
decomposition of $Q\supset P$ for neutral species. In this, the Higgs mass also
acts like a threshold energy above which the proper length that a particle of
energy $E_{\rm p}$ travels through 3-space is reduced by a factor of
$m_{\rm H}/E_{\rm p}$ due to the topological identifications on $L(T^3)$.

\section{The Equation of Motion}

\subsection{Derivation}

The interactions present in the equation of motion {\it must} follow directly
from the topological structure of $Q$ and $L(T^3)$ {\it if} one accepts the
thought experiment. One demands on the left hand side a single index equation
because there are four homotopically inequivalent paths on $L(T^3)$; cubic
interactions since every $S^1$ loop is attached
to a handle and a junction; and scalar quadratic interactions and scalar second
derivatives due to individual loops.
The right hand side is zero because $Q$ and $L(T^3)$ are compact.
This yields for the complex four vector $q_\lambda$
$$q^\mu [\partial^\nu q_\mu ,\partial_\nu q_\lambda ]=0,\eqno(8a)$$
where only the commutator form satisfies the right hand side for solutions of
the Klein-Gordon equation, i.e.\ the low energy limit. One finds
$$q^\mu [q_\lambda\Box q_\mu -q_\mu\Box q_\lambda ]=0.\eqno(8b)$$

The square of the absolute value of the wave function,
$q^\mu q_\mu\equiv \delta^{\mu\nu}q^*_\nu q_\mu$, assures a positive definite
inner product and a well defined probability distribution. The individual
components of $q_\lambda$ yield probability distributions for each of
the four homotopically distinct paths through $L(T^3)$. Since $Q$ and $L(T^3)$
are nuclear, the theory is manifestly Lorentz invariant with a Minkowski metric
$\eta^{\mu\nu}$, $\partial^\mu =\eta^{\mu\nu}\partial_\nu$, and $\Box$ the
d'Alambertian.

Finally, unlike the three-tori, the mini
black holes couple directly to the matter degrees of freedom through the
process of Hawking radiation. Therefore, handles can be viewed as being
continuously created by and destroyed (see the thought experiment).
These quantum perturbations in the local number of handles lead to the
generation of an additional field. This field is envisaged to reflect phase
changes (the topology of $Q$ is not altered) in the wave amplitudes
$q_\lambda$, flowing through $L(T^3)$. The fundamental object to solve for on
$Q$ is therefore $\Omega_\lambda\equiv {\rm e}^{2\pi i\phi}q_\lambda$, with
$\phi$ a real function of time and position. This phase transformation leads
to the equation of motion on $Q$
$$4\pi i\partial_\nu\phi [(\partial^\nu q_\lambda )q^\mu q_\mu -(\partial^\nu
q_\mu )q^\mu q_\lambda ]=q_\lambda q^\mu\Box q_\mu -q^\mu q_\mu \Box
q_\lambda ,\eqno(9)$$
with an additional scalar constraint
$$q^\mu q_\mu =1,\eqno(10)$$
which becomes void on $L(T^3)$. Equation (10) is invariant under the phase
transformation and the $\mu$ summation reflects the fact that it is an
observable. The scalar constraint thus signifies
that, due to the continuous creation and destruction of handle
manifolds, it is possible to travel from one point along a homotopic path to
any other point along a different homotopic path in $Q$. Therefore, from the
perspective of the wave amplitudes, any point in $Q$ becomes indistinguishable
from any other, while the topology of the thought experiment persists. In other
words, $Q$, unlike $L(T^3)$, contains no scales, expressable as functions
$\ell_{\rm Planck}()$ and $m_{\rm Planck}()$, to distinguish one set of points,
internal to an $S^2$ horizon, from any other.

Because the evolution of $\phi$ is driven by the handles, and $L(T^3)$ has a
topology which is distinct from $Q$, it follows that the condition
$\partial_\nu\phi =0$, which reduces (9) to (8) and has been referred to as
the evaporation of the handles so far, corresponds to a quantum
transition. The numerical value of the field $\phi$, or rather $\Delta\phi$
since (9) is invariant under the global transformation
$\phi\rightarrow\phi +\beta$, must
correspond to a constant, not necessarily zero, Lorentz invariant vacuum
expectation value, i.e.\ $<0|\Phi |0>$, under the junction potential $V$ on
$L(T^3)$. A non-zero vacuum expectation value of $\Phi$ requires
$\mu^2<0$ in $V$ and can lead to spontaneous symmetry breaking.
The additional scalar $\phi$ can thus induce a Higgs field $\Phi$.
This is a doublet due to the quantization rule for $|\Delta\phi |$ discussed
immediately below.

\subsection{Interpretation}

\subsubsection{Evolution and Inflation}

The $Z_3$ group of the handle triplet yields a discrete spectrum
$\Delta\phi =\pm 1/3,0$, i.e.\ the topology of $Q$ determines the quantization
conditions. The rule $|\Delta\phi |=1/3$, applied to the solution space of (9)
and (10), is then tantamount to the identification of inflationary solutions
because a non-zero value of $\Delta\phi$ implies $\mu^2<0$ in $V$ and hence the
availability of a vacuum energy.

Below it is shown that the zero point vacuum energy is uniquely determined by
the topology of $Q$, so that the equation of motion knows about it. The
possible durations $\Delta t$ over which inflation occurs in each identified
solution are then determined by $\Delta t=|t-t'|$ for all space-time points
that satisfy $|\Delta\phi |=|\phi (x,y,z,t)-\phi (x',y',z',t')|=1/3$, where
$t$ is a formal time coordinate which only has a physical meaning for the size,
$\sim {\rm e}^{\Delta t/t_{\rm Planck}}$, evolution of $L(T^3)$. Note here that
$Q$ is characterized by the dimensional numbers $\ell_{\rm Planck}$ and
$m_{\rm Planck}$, but that it has no fixed physical dimensions. Indeed, it is a
ground state which defines a solution space of initial conditions for $L(T^3)$.

The initial conditions at $t=0$ for
the solutions of (9) can be taken as $q_\lambda (0)={\rm cst}$,
derivatives $\partial_t q_\lambda (0)=1$ in Planck units, and $\phi (0)=0$.
To follow the evolution of the wave function after the handles have
evaporated\footnote{If merging contributes significantly, i.e.\ there are very
strong peaks in $q^\mu q_\mu$, then a lot of primordial black holes may be
formed.}, one should solve equation (8) with the end solution of (9) as initial
conditions. During this phase, the characteristic amplitude of the
fluctuations is $\delta\rho /\rho (Q)$ as computed above because
all the original 23 degrees of freedom of $Q$ (later to become particles) are
above the GUT unification scale (see its computation below). Once the GUT is
broken at some energy, the Einstein equation describes the later time
evolution of the mass-energy distribution (the expectation value of
$q^\mu q_\mu$), as the universe expands. The fact that general relativity does
not constrain the topology of space-time thus appears to follow from the fact
that it is only valid if one can ignore the topology of space-time.
Nevertheless, Einstein gravity is still a part of TD through the presence of
the handles.

\subsubsection{Multiple Connectedness}

Equations (8), (9) and (10)
describe the quantum-mechanical interactions of mass-energy in full,
i.e.\ including quantum gravity, without any need to know the specific
properties of the particles in the field theory. The boundary
conditions for the solutions to these equations are {\it topological} and
follow from the cyclic properties of $L(T^3)$.
The topology of $L(T^3)$ requires the solutions $O_\lambda (x,y,z,t)$
to be periodic on scales $L_i=n_i\ell_{\rm Planck}$ for positive
integers $n_i$, $i=1..3$, and at every time $t$,
$$O_\lambda (x,y,z,t)=O_\lambda (x+L_1,y+L_2,z+L_3,t).\eqno(11)$$
Note that these are {\it identifications} between space-time points
which need {\it not} be an infinitesimal distance apart.
The specific realizations of these identifications in a field theory will be a
major topic in subsequent sections.

\subsection{Rest Masses}

A question which can be addressed through solutions to (8) and (9) is the
nature of the statistics of {\it mass-energy} fluctuations during the
radiation-dominated era, after the GUT is broken. Furthermore, the solutions of
(8) on $L(T^3)$, with the specific volume $\delta\rho /\rho (T^3)=7^{-3}$, then
provide the possible probability distributions for the {\it rest masses} of
particles. That is, for some ground state mass $m_0$, one has
$m=m_0[1+q^\mu q_\mu\delta\rho /\rho (T^3)]$, where $q^\mu q_\mu$ must be
normalized to unity on the unit cube, and with a similar expression for the
mass-energy in terms of $\delta\rho /\rho (Q)$.

It follows that the rest masses of particles must be a function of
position {\it if} the universe was not perfectly homogeneous at the GUT
transition. The reason for this conclusion is a direct consequence of the
thought experiment in which {\it units of} mass enter through Plankian black
holes that are not necessarily well localized in space due to the topological
identifications supported by the underlying $L(T^3)$.

\section{Predictions for the Standard Model}

In order to connect with the low energy Universe, one should relate TD to
the standard model, and assess whether it is capable to reproduce the
masses of particles and the values of coupling constants.
As far as the former are concerned, it will be shown that the expectation
values of particle rest masses are topological, i.e.\ they are elements of the
rational numbers multiplied by $m_{\rm Planck}$. The probability distribution
around this expectation value is as discussed immediately above in \S 4.3.

\subsection{Unification Energies}

When the handles evaporate, the value of $|\Delta\phi |$ can be $1/3$ which
breaks the $SU(5)$ symmetry through $V$. Since $Q$ is the ground state, the GUT
energy scale should correspond to $1/3$ ($Z_3$ introduces three branches) times
the energy per degree of freedom on $Q$ per three-torus.
One finds $M_{\rm GUT}=1/3m_{\rm Planck}/2N_Q= 8.8\times 10^{16}$ GeV,
with $m_{\rm Planck}=1/\surd G=1.22\times 10^{19}$ GeV in units where
$\hbar =c=1$. Above it was shown that the latent heat per $T^3$ associated with
handle evaporation, equals $h=9/46m_{\rm Planck}$. Therefore, one finds
$M_{\rm A}=h/3= 8.0\times 10^{17}$ GeV, for the energy at which $L(T^3)$
emerges\footnote{Note that on $L(T^3)$ these energies {\it per degree of
freedom} are a factor of 7 smaller than $M_{\rm GUT}$ and $M_{\rm A}$.}.

\subsection{The Absolute Scale of the Mass Ground State}

The 23 degrees of freedom of $Q$ define
$X_Q=23!$ different configurations. The mass
of the particle ground state is thus $m_{\rm Planck}/X_Q$. The particle should
be charged because $Q$ contains a $U(1)$ sector. It follows that
$$m_{\rm e}^0=m_{\rm Planck}/X_Q= 0.47\quad {\rm MeV},\eqno(12a)$$
determines the electron mass ground state.

For the submanifold $P$ one has $X_P=14!$, which fixes the neutral
ground state on $L(T^3)$. Because a ``neutrino'' has no charge, it cannot be
distinguished on $P$, unlike the charged leptons which couple to the $U(1)$
sector on the junction of $Q$. The total number of configurations is now
$X_QX_P$. The mass of the neutral (electron neutrino) ground state thus follows
from
$$m_{\nu_{\rm e}}^0=2m_{\rm Planck}/(X_QX_P)= 1.08\times 10^{-5}\quad
{\rm eV},\eqno(12b)$$
where the factor two results because the neutral neutrino cannot be
distinguised on either three-torus under the transition $Q\rightarrow L(T^3)$.

\subsection{Corrections to the Mass Ground State}

If $\delta\rho /\rho (Q)=23^{-3}$ is the specific volume of a probability
distribution on $Q$, then one can ask with what accuracy
${\cal A}$ the properties of $P\subset Q$ can be determined, given that
$\delta\rho /\rho (P)=14^{-3}$. The uncertainty relation
yields $\delta\rho /\rho (Q)={\cal A}\delta\rho /\rho (P)$. This question
is relevant to the particle mass in the
charged and neutral ground state since the handles occupy only a part of the
total number of degrees of freedom on $Q$. One finds ${\cal A}=(14/23)^3=0.23$,
which reflects an upward shift in the mass because the finite accuracy
${\cal A}$ implies that less information is needed to describe the system,
i.e.\ an effectively larger ``measuring rod''.
From the $Z_3$ symmetry of $\Theta$ (three branches), one finds
that the magnitude of the shift is ${\cal A}/3$ which yields
$$m_{\rm e}=(1+{\cal A}/3)m^0_{\rm e}=0.51 \quad {\rm MeV}.\eqno(13a)$$
This is in excellent agreement (error $<0.7$\%) with the measured value of
$0.511$ MeV, and lends support to the notion that the 23 degrees of freedom of
$Q$ are truly fundamental. Analogously, one finds that
$$m_{\nu_{\rm e}}=(1+{\cal A}/3)m_{\nu_{\rm e}}^0=1.16\times 10^{-5}\quad
{\rm eV}.\eqno(13b)$$
The fact that the electron neutrino has a mass is a unique prediction of
the model.

\subsection{The Masses of Other Elementary Particles}

What follows below is an, at times tedious, enumeration of the various
topologically distinct configurations of $Q$ and $L(T^3)$. In the topological
approach advocated here, one should view the expectation value rest mass of
a particle as an energy level in an atom which is fixed by the
ambient degrees of freedom.

The $SU(3)$ gauge potential of QCD is realized
on the junctions of the lattice, and therefore acts on pairs of quarks.
Since each three-torus in $L(T^3)$ has six neighbors, there are $6!$
physically distinct orderings for the adjacent quarks. It follows that the
doublet mass ground state is $m_{\rm u,d}=6!m^0_{\rm e}=340$ MeV, which is
split by an amount $\delta m={\cal A}/3m_{\rm u,d}$ partitioned over $P$.
The s, c, b, and t excited mass
states represent four topologically distinct configurations of the lattice,
which yield mass multiplication factors given by the ratios of degeneracies
$\delta_{\rm q}/\delta_{\rm u,d}$, with $\delta_{\rm u,d} =2$. One
finds $\delta_{\rm s}=3$ and $\delta_{\rm c}=2\delta_{\rm s}$ because there
are three fermionic states (pairs of $S^1$ loops) supported by any three-torus,
and twice that many by the doublet. It further follows that
$\delta_{\rm b}=\delta_{\rm s}^3$ due to their possible combinations into
triplets. Finally, $\delta_{\rm t}=4/3\delta_{\rm s}^6$ for the doublet, where
there are 4 homotopically inequivalent paths between the two three-tori, but 3
of these are associated with either $T^3$ itself (the 1 is of course $S^3$).
From these numbers one finds rest masses for s, c, b, and t given by
$m_{\rm s}=510$ MeV, $m_{\rm c}=1.02$ GeV, $m_{\rm b}=4.59$ GeV, and
$m^0_{\rm t}=165.3$ GeV, respectively. The observed top quark mass is
$m_{\rm t}=(1+{\cal A}/3)m^0_{\rm t}$, because it represents a topological
state on $L(T^3)$ that fixes one of the six $U(1)$ junctions around a
three-torus. The derived quark masses are in good agreement with experiment.

The excited lepton states scale with $M_{\rm A}^{-1}=138/9$ in Planck units,
because the emergence of $L(T^3)$ yields the global electromagnetic field for
the $\Theta$ sector. The ratios
of the excited state masses to the ground state must be double and triple
powers of $M_{\rm A}$, because the equation of motion (8) supports cubic
self-interactions, with a quadratic scalar term. This yields masses for the
muon and tau of $m^0_\mu =119.4$ MeV and $m^0_\tau =1831$ MeV, respectively,
up to topological mass corrections, in good, 13\% and 3\%, agreement with
observations. For their respective neutrinos one finds
$m^0_{\nu_\mu}=2.74\times 10^{-3}$ eV and $m^0_{\nu_\tau}=0.042$ eV.

Analogous to the ground state masses, there is a purely topological correction
to these excited state rest masses due to the structure of $Q$ and $L(T^3)$.
The transition $Q\rightarrow L(T^3)$, where $Q$ contains two three-tori, yields
an intrinsic accuracy ${\cal B}=(7/14)^3$. This number signifies the extra
information regarding the $\mu$ state which is needed for the transition to the
excited state, i.e.\ an effecticely smaller ``measuring rod''. The sign of
this mass correction is thus always negative, $-{\cal B}m_\mu$.
One finds $1/4$ this number for the $\tau$ because the latter
is a third order state, and can have any of the four paths through
$L(T^3)$ attached to a (quadratic) $S^1$ loop. The same corrections
apply to the associated neutrinos, but not to the quarks since their masses
derive from the global topological properties of $L(T^3)$ alone. Hence the use
of $m^0_{\rm e}$ in the doublet quark ground state above.
Once applied, these corrections yield approximately $\sim 7^{-3}$ agreement
with experimental results,
$m_\mu =104.5$ Mev, $m_\tau =1774$ MeV, $m_{\nu_\mu}=2.39\times 10^{-3}$ eV,
$m_{\nu_\tau}=0.041$ eV. The neutrino results are consistent with
SuperKamiokande findings\cite{6}. In
Section 8 the electroweak mixing angle and fine
structure constant are computed, and the topological masses of the vector
bosons are presented there.

Furthermore, the mass of the Higgs
boson follows from $\lambda^{-1}=7$, the number of degrees of freedom of an
individual three-torus in $L(T^3)$ on which the self-interaction potential $V$
lives, and $m_{\rm H}=\surd (2v^2\lambda )$ with the Fermi
coupling constant $G=1.16632\times 10^{-5}$ GeV$^{-2}=v^{-2}/\surd 2$ (see
Section 8 for its predicted magnitude). This yields a value for the Higgs boson
mass $m_{\rm H}=131.6$ GeV, consistent with current experimental limits.
It should be pointed out that the value of $m_{\rm H}$ is independent of any
supersymmetry. In fact, the latter does not enter the discussion other than
that $T^3$ happens to have a holonomy $H=1$, and $L(T^3)$ can support both
$O(4)$ and $SU(2)$ (see Section 8).

Note that the probability distribution $q^\mu q_\mu$ for the rest masses of
particles is the same for all particle species, but that collapse
proceeds independently for the ground and excited states because they are
associated with topologically distinct configurations.
Given that our Galactic environment is the result of a primordial density peak,
it is indeed to be expected that most of the expectation values computed above
are on the low side by a few times $7^{-3}$.

\section{Spatial Variations in Inertia}

\subsection{Non-Gaussianity}

The equation of motion (8) on $L(T^3)$ is solved by solutions to the linear
Klein-Gordon equations
$$\Box q_\lambda =-m^2q_\lambda,\eqno(14)$$
where the solution space is restricted by (11), and (14) can be rewritten as a
general Proca equation for $m^2\not= 0$ since the latter then yields
$\partial^\mu q_\mu =0$. The solutions to (14) are
taken here to be quadruple travelling wave forms for $m^2=0$
$$q_\lambda =f_\lambda (l_1x+l_2y+l_3z-t)+f_\lambda (l_1x+l_2y+l_3z+t),
\eqno(15)$$
where the direction cosines $l_i$ satisfy
$$l_1^2+l_2^2+l_3^2=1,\eqno(16a)$$
the functions $f_\lambda$ have the standard forms for this wave
equation limit, and the boundary conditions (11) further require that
$$l^iL_i=g^{ij}l_jL_i=0,\eqno(16b)$$
with $g^{ij}$ the three-metric which relates the topological boundary
conditions to the geometry of the Universe. The
coefficients $L_i$ determine the physical shape of the wave forms.
It is the superposition $S$ of this set of solutions over all possible $L_i$,
and the $l_i$ they allow, which determines the inertia of material particles
in the Universe. Note that the functional forms in normalized coordinates of
the $f_\lambda$ generally will be different, depending on the solutions to (9).

The superposition of the $q_\lambda (l_i[L_i])$ through the dependent
coefficients $l_i$ need not be Gaussian. In fact, an investigation of the
constraints for $g^{ij}=A(t)\delta^{ij}$ with a scalar $A$ shows that $S$ has
non-Gaussian features because very anisotropic wave packages (lines and sheets)
are part of the solution space, and move almost orthogonally to their major
axis/axes, {\it if} one assumes that there are no preferred identifications
between various three-tori. As all these sheets and lines are
superposed, one creates small scale and large amplitude fluctuations in both
the mass-energy and particle rest masses. From the topological considerations
above it further follows that the isotropic perturbations exhibit decoherence
that is linear in scale.

\subsection{Primordial Fluctuations and Matter/Anti-Matter Inequality}

Observe with respect to the different amplitudes for mass-energy and
rest mass fluctuations an interesting transition at decoupling, when the
photon energy density no longer dominates that of the matter. The respective
amplitudes for mass-energy and rest mass perturbations are given by the
specific volumes $23^{-3}$ and $7^{-3}$. These differ by a factor of
approximately 35. To satisfy the $\delta\rho /\rho (Q)$ constraint for the
ambient mass-energy fluctuations, one requires spatial variations in the number
density of massive particles on the order of $7^{-3}$. At the epoch of
decoupling these number density fluctuations induce particle diffusion
from number overdense to number underdense regions, and thus selectively
increase (bias) the amplitudes of mass overdense regions.

The amount of matter/anti-matter asymmetry is determined by the local $7\sigma$
functional values of $q^\mu q_\mu$ at the time of GUT symmetry breaking. That
is, $Q$ supports both particle and anti-particle sectors but their relative
proportion on $L(T^3)$ is determined by the collapse of $q^\lambda$ on the
sevenfold three-torus.

\section{Black Holes}

For black holes the solutions to (14) are taken to be generic spherical waves
$$q'_\lambda =\Sigma_{k,s}(2\omega_k v)^{-1/2}\epsilon_{\lambda s}(k)[a_s(k){\rm
 e}^{-i(-kr+
\omega_k t)}+b^\dagger_s(k){\rm e}^{i(-kr+\omega_k t)}],\quad s=1,2,3,\eqno(17)$$
with orthonormal polarization vectors $\epsilon_s$, the Lorentz condition,
wave vector $k$, proper volume
$v$, $\omega_k^2=m^2+k^2$, the standard interpretation in terms of annihilation
and creation operators, analogous expressions for photons with $m^2=0$, and the
index $\lambda$ for the four homotopically distinct paths on $L(T^3)$. For
the $S^1\times S^2$ topology of the handle one finds the boundary conditions,
which are again topological identifications,
$$q'_\lambda (r=0,t)=q'_\lambda (r=a',t),\eqno(18)$$
in spherical coordinates, with $a'$ the Schwarzschild radius in Planck units.
Therefore, all the mass-energy crashing into the singularity is represented on
the event horizon. Note that the localization of the black hole's information
like this can only be established through the possible topological
identifications in $L(T^3)$, and is in fact required by the existence
of event horizons of arbitrary size which Nature must accommodate.
That is, the thought experiment is valid for black holes of any size and a
particle entering the event horizon must hit the singularity.
Furthermore, one has the parameterization of the wave vectors
$$k_n={{2\pi}\over{a'}}n, \quad n=\pm 1,\pm 2,...,\pm a',\eqno(19)$$
along the $S^1$ of the handle.

\subsection{The Cosmological Constant: Geometry and Topology}

Note first the difference between the left and the right hand
side of the Einstein equation. Section 10 will deal with the right, the zero
point energy associated with the particle vacuum, and this section with the
left, curvature associated with a non-trivial, but massive, topology (which is
not fixed by general relativity). Clearly, this distinction must be arbitrary
at some fundamental level, as the following discussion and Section 10 will
examplify.

The boundary conditions are half the solution to the equations of motion. The
key to understanding $\Lambda_{\rm BH}$ is precisely in these, as follows.
The solutions (17) must match the rest of the Universe outside of the event
horizon if black hole
evaporation is to occur on thermodynamic grounds at some point in
time, i.e.\ (18) only describes the black hole as a closed system. Let the
collapse be denoted by $\{a'_i\}$ with $i=1$ corresponding to $r=0$. The length
of any body as a function of proper time in Planck units
$a''_i\propto(\tau [a'_i]-\tau [a'_1])^{-1/3}$ reflects the matching condition
for the surface $\tau [a'_i]$ which characterizes a configuration
$i$. Because $V\propto a_i''^{-3}$, one finds a peak, the singularity,
in $q^\mu q_\mu$ for particles represented by (17) near $i=1$. Note that the
mass-energy {\it and} particle rest masses follow this peak. It is therefore
possible, as mentioned above, to concentrate all the information on the black
hole horizon through (18), while using the singularity as the locus of the
mass-energy. Since the resulting probability distributions are normalized, only
the $i=1$ configuration gives a non-zero contribution under
$\lim_{V\to 0}\int_V(q^\mu q_\mu )_i$.

Furthermore, $a''_i$, like $a'$, is a geometric quantity which incorporates the
shape of space-time. It is the function $a''_i$, similar to (16b), that
provides the relationship between the topology and the geometry of space-time.
That is, information is contained in the wave modes (17), whose absolute
extend, like the global scale invariance of (16b), derives from space-time
geometry. It follows that the modes associated with the black hole singularity,
$i=1$, extend over distances of size $a'^2\ell_{\rm Planck}$. Clearly, such a
conclusion depends crucially on the existence of the topological identification
(18) on $L(T^3)$ as discussed above.

In conjunction, the rest masses diverge at $r=0$.
This allows Planck mass excitations to be generated throughout the Universe,
and render the singularity finite by associating it with any point beyond the
horizon, for a Planck time, while preserving the handle topology.
These associations can be viewed as mini black hole pairs (both signs of $n$)
which constitute the left hand side cosmological constant\cite{3}.

The black hole singularity is thus realized
as a polarization of the vacuum by the matter degrees of freedom. Although
this result might appear counterintuitive, the introduction of a topological
origin for the inertia of massive particles leads directly to 't Hooft's
suggestion that black holes are a natural extension of particles,
and hence to a link between topology and the properties of the vacuum. When the
black hole has evaporated the singularity vanishes into that same vacuum.
Finally, the contribution $\Lambda_{\rm BH}$ to the cosmological constant is
given by the number density of black holes times $2m_{\rm Planck}$.

\subsection{Information, States and Resonances}

Note that solutions to the Dirac equation also solve (14). In fact, the four
components of $q_\lambda$, a topological property, facilitate a match
for the four Dirac spinor components, given the $SL(2;C)$ spin structure on the
four-manifold bounded by a nuclear prime manifold $T^3$.
Thus, the mass, charge and spin of any particle can be encoded in
the wave modes, and black hole
evaporation can proceed unitarily in this sense. Nevertheless, one will
need the complete evaporation history for the reconstruction of the data
\footnote{Suggestions by M.\ Bremer on these matters are greatfully
acknowledged.}.

It is straightforward to predict the number of black hole states $N_{\rm BH}$.
There are $2i$ wave modes (quanta) at a given $a'_i$, and 1 in 4 of these
contains relevant information since $\lambda =1..4$ allows the construction of
every double, triple and quadruple wave number of a mode through
multiplication. The state space of a black hole $a'$ then consists of all
configurations formed by the sum of its $a'_i$. This yields, for an integration
over $4\pi$ steradians, the entropy $S_{\rm BH}$
$$S_{\rm BH}/4\pi =\thinspace ^7{\rm log}N_{\rm BH}=
{{1}\over{4}}\Sigma^{a'}_{i=1}2i=(a'^2+a')/4,\eqno(20)$$
in Planck units, with 7 the natural base number for $L(T^3)$, i.e.\ the
smallest unit of information is $ln 7$. One has the quantization condition
that $(a'^2+a')/4$ be integer valued, which gives $a'=3$ as the smallest
solution with $N_{\rm BH}=7^3$. For large $a'$ one has the semi-classical
result $S_{\rm BH}=A/4=\pi (a'^2+a^2)$, with $A$ the area and $a$ the angular
momentum per unit mass. A comparison shows that only that part of the surface
area associated with the event horizon, $a'^2=[M+\surd (M^2-Q^2-a^2)]^2$ for a
charge $Q$, contributes to the number of quantum states (20). Therefore, the
ergosphere constitutes a region for interaction with the internal states (which
are quadrupally degenerate), that adds to the total entropy but whose
information content is intrinsically macroscopic. Thus, $\pi a^2$ is
effectively an integration constant.

The black hole temperature $T$, associated with $A$ through the irreducible
mass $M^{\rm ir}=(A/16\pi)^{1/2}$, determines
the internal structure of the black hole, and is relevant to its statistical
properties on $L(T^3)$. That is, the black hole emits a
spectrum constrained by the {\it summed}, because of (18), incoming wave
function, where the temperature $T$ determines the probability for re-emission
through ${\rm exp}[-E_{\rm p}/kT]$. The energy $E_{\rm p}$ is the particle's
energy when it originally entered the black hole.

Furthermore, absorption of neutrinos by the black hole will lead to stimulated
emission because of the structure of $Q\supset P$ for neutral elementary
particles. In this, the total mass (adding $\surd (m_{\nu}m_{\rm H})$ per
absorbed neutrino $\nu$), spin and angular momentum accreted in neutrinos since
the formation of the horizon form a global constraint for the emission.

Finally, the
purely topological properties of the black hole imply the existence of
resonances, at wave numbers $k_n$ and with a (constant) width of $7^{-3}k_n/n$,
where the black hole behaves like a wormhole for any ingoing path whose
starting point is locked in phase (to better than about one part in $7^3$) with
some end point through the paired emission of lines at resonance frequencies.
The end point then induces an outgoing $S^2$ surface of radius $a'$ around
itself. The same holds for the reverse path, where the exit kinetic energies of
the transferred modes are fixed by the work associated with the trajectories
upto the starting point from infinity and from the event horizon, respectively.
In fact, the only defining characteristic of these resonances is information.
Therefore, a wormhole is created if a resonance is achieved with any,
arbitrarily far away, black hole. Note that the overlap of many
resonances, e.g., due to multiple reflections of the resonant signal combined
with Doppler shifts and black hole mass ranges, can lead,
among other things, to localized time lapses of the order of $<a'>/c$ divided
by the probability $p$ for coherence. In this, $<>$ denotes some Universal
average and $p$ is derived by using the mass energy / particle rest mass
distribution $q^\mu q_\mu$, at the time of GUT breaking, to determine the
probability distribution for fluctuations with respect to the vacuum to occur
{\it anywhere}.

\subsection{The Cosmological Constant: Mass}

It should be noted that any particle crossing the
event horizon attains a rest mass given by its vacuum value $m_0$. It is this
vacuum value that is relevant to the mass of the black hole. As such, a region
with some value of $q^\mu q_\mu$ for the rest masses of particles therein, will
exhibit a decrease in its mass when enclosed by an event horizon.
Hence, the presence of large amplitudes in $q^\mu q_\mu$ and their subsequent
collapse into black holes can cause an accelaration in the expansion of the
Universe.

\section{The Lattice Structure of Space-Time}

\subsection{QCD}

It is easy to see that the $U(1)\sim O(2)$ group on the {\it junctions} of the
$L(T^3)$ lattice, leads to {\it sub}-structures
given by two-dimensional periodic Ising models.
One also finds that this system as a whole is
{\it frustrated}\footnote{The author is very greatful to B.\ Canals for
discussions on this point.}, which is a fascinating realization given the
importance of the strong interaction in the Universe.
The very presence of frustration in fact requires the existence of collective
excitations on the Ising driven lattice, and as such provides a possible
physical basis for the concept of confinement under the QCD gauge group.

This result suggests that it is worthwhile to explore the rather obvious
fact that $SU(3)\supset U(1)^3$. Consider the nearest neighbor Hamiltonian
$${\cal H}=\kappa\Sigma_{i,j;i\not= j} [S_i^kS'_{ik}]\cdot [S_j^kS'_{jk}]
=\kappa\Sigma_{i,j;i\not= j}\sigma_i(\{ I\} )\cdot\sigma_j(\{ I'\} ),
\eqno(21)$$
where $S_k=S'_k$ is excluded if it holds for all $k$ up to a global sign,
$k=1..3$ labels the three $U(1)$ ``color'' junction variables $S_k=\pm 1$.
Furthermore,
$\sigma =\pm 1$ and the symbol $\{ I\}$ denotes a set of internal indices
discussed below. Periodic boundary conditions need to be applied if the
Universe is closed. The form (21) is invariant under transformations
$S_k''=MS_k$ with $M\in SU(3)$.

The 8 gluons in QCD are a direct consequence of the number of different $S_k$.
It is easy to see that there are $2^3=8$ triplets $S_k$. The contraction of
various $S_k$ yields, when grouped together in doublets differing only by an
overall sign, 12 possibilities, i.e.\ 6 quarks with their anti-particles.
One always finds $S^kS'_k=\pm 1$, yielding a true spin lattice and hence the
second equality in (21). But, due to the vector nature of the $S_k$, every
$S^kS'_k=\pm 1$ must carry {\it internal} indices $\{ I\}$ which indicate
quark type and color.

Thus, the internal variables of the $\sigma$ living on the lattice junctions
provide discrete degrees of freedom through weak interactions (flavor change),
and photon and gluon production. The latter two are special in that the
electromagnetic field allows quark/anti-quark pair annihilation, which is
equivalent to the introduction of a (temporary) lattice defect. The gluons
facilitate quark/anti-quark pair flavor change and pair production, as well as
quark color conversion. The sole emission of a gluon by a quark leads to the
creation of a quark/anti-quark pair at an empty couple of lattice sites. All
in all, the indices $\{ I\}$ are amenable to specific rules while the global
equilibrium is determined by the (frustrated) behavior of the $\sigma$ on the
lattice junctions. In any case, it appears that powerful Monte Carlo techniques
developed for condensed matter physics on frustrated lattices should be
applicable to QCD physics.

Finally, the coupling constant $\kappa$, at the energy scale of the Higgs
boson, is given by $\kappa =9^{-1}(1+{\cal A}/3)=0.1195$, and it reflects the
degrees of freedom of the $\Theta$ sector, which are not part of QCD for any
realization of $Q\supset P$. This value is in good agreement with the current
world average (at the mass of the $Z$ boson)
$\kappa_0=0.1188\pm 0.0018$\cite{7}.

\subsection{Other Coupling Constants}

\subsubsection{The Fine Structure Constant}

The GUT energy scale $M_{\rm GUT}=1/138$ in Planck units should define the QED
coupling constant since it reflects the emergence of the individual particles
from the ground state $Q$, that can travel along the $U(1)$ junctions and
interact with the global photon degrees of freedom. The measured
fine structure constant $\alpha_0=1/137.0359895$
agrees with $K$ to 0.7\%, where the discrepancy is due to the presence of a
non-trivial space-time topology. The manifold $Q$ supports
$\eta_Q=(\delta\rho /\rho (Q))^{-1}+1=23^3+1$ dynamical degrees of freedom
through the equation of motion, where the 1 is for its time
evolution. For $L(T^3)$ one has $(\delta\rho /\rho (T^3))^{-1}+1=7^3+1$
dynamical degrees of freedom, but the lattice has a four-fold, $e_{\rm M}=4$,
degeneracy, yielding $\eta_L=e_{\rm M}^{-1}(7^3+1)=86$
(indeed an integer). This yields
$$\alpha '=M_{\rm GUT}(\eta_Q-1)/(\eta_Q-\eta_L)=1/137.0359168\eqno(22a)$$
for the
fine structure constant, corrected for those degrees of freedom
associated with the excited state character of the Universe, which are not
confined to the ground state at any point in time.

There is another
correction associated with the fact that the total system, ground state plus
excited state, has $\eta_{\rm t}=e_{\rm M}(\eta_Q-\eta_L)\eta_L$
effective degrees of freedom, but quantization, fixed by the {\it constants}
$\ell_{\rm Planck}$ and $m_{\rm Planck}$, removes two of
these to define a physical theory.
This gives a correction factor $f=(\eta_{\rm t}-2)/\eta_{\rm t}$.
One finds
$$\alpha =f\alpha '=1/137.0359828,\eqno(22b)$$
with a relative accuracy of
$4.9\times 10^{-8}$, consistent with the experimental $1\sigma$ uncertainty in
$\alpha_0$ of $4.5\times 10^{-8}$. The numerical value of $\alpha$ then fixes
the unit of charge $e$. Note that the constancy of the speed of light is an
automatic consequence of nuclearity of the three-torus on $L(T^3)$, and hence
one does not have to correct for it.
In closing it should be noted that $\alpha$ is a global parameter (contrary to
$\kappa$) in that it contains information on the joint characteristics of $Q$
and $L(T^3)$, i.e., on both the number of unrestrained physical parameters in
the theory and on its number of dynamical degrees of freedom. Its running value
at the energy scale of the Higgs boson is $(1+{\cal A}/3)\alpha$.

\subsubsection{The Weak Coupling Constant}

The weak coupling constant follows from
$g^2_{\rm W}/4\pi =4/7\cdot 86/85\alpha$ because the ratio of these coupling
constants for the electroweak sector must reflect the respective number
of degrees of freedom of the lattice ($e_{\rm M}=4$ paths between any two
points, long range) and of $T^3$ (multiplicity of seven, short range),
and is modified for the photon degree of freedom. With the standard definition
of the weak mixing angle ${\rm sin}^2\theta '_{\rm W}=\zeta$, this yields
$\zeta =8^{-1}[7/4\cdot 85/86]=0.216$, where $\zeta$ should be viewed as
including all higher order quantum corrections.
One thus finds the masses for the vector bosons
$m_{\rm W}={\rm sin}^{-1}\theta '_{\rm W}(\pi\alpha /G\surd 2)^{1/2}=80.18$ GeV
and
$m_{\rm Z}=2{\rm sin}^{-1}2\theta '_{\rm W}(\pi\alpha /G\surd 2)^{1/2}=90.56$
GeV, consistent with experimental limits and the $7^{-3}$ intrinsic uncertainty
associated with the collapse of their rest mass wave functions.

The effective value of the weak mixing angle, at the energy scale of the Higgs
boson is ${\rm sin}^2\theta_{\rm W}=(1+{\cal A}/3)\zeta =0.23246$, corrected
for the intrinsic accuracy ${\cal A}/3$ of $Q$. This distinction is a
result of the fact that the vector bosons are represented in the ground state,
while they derive their masses from the non-zero vacuum expectation value of
the Higgs field under $V$ on $L(T^3)$.
The value for ${\rm sin}^2\theta_{\rm W}$ is in poor agreement
with the overall mean experimental value of $0.23148\pm 0.00021$, but in good
agreement with the LEP average of $0.23196\pm 0.00028$ (both at the mass of the
$Z$ boson).

\subsubsection{The Fermi Coupling Constant}

The Fermi coupling constant $G=\surd 2 (g_{\rm W}/m_{\rm W})^2$ is
fixed through
$G^{-1/2}=-m_{\rm H}+\Sigma_i m_i=293.2$ GeV, i.e.\ by the sum of
all the particles represented in the ground state minus the
Higgs boson. The latter reflects spontaneous symmetry breaking, which is
associated with the excitation of the ground state. The value
$G=1.163\times 10^{-5}$ GeV$^{-2}$ is consistent with the experimental limit of
$G_0=(1.16632\pm 0.00002)\times 10^{-5}$ GeV$^{-2}$, given the intrinsic
$7^{-3}$ uncertainty in the collapsed rest mass wave functions. In fact, since
some of the mass terms on the right hand side depend on the value of $G_0$,
{\it a disagreement would have ruled out TD theory immediately}.

\section{CP Violation and Time Delays}

There is one more degree of freedom on $L(T^3)$. The lattice junction of a
pair of three-tori can support a topological identification
along the time axis. The
energy $F$, corrected for ${\cal A}/3$, of the excitation is
$$F=(1+{\cal A}/3){{m_{\rm Planck}-H}\over{X_QX_P}}=3.546\times 10^{-6}
\quad {\rm eV.}\eqno(23)$$
The denominator reflects the total number of neutral configurations on $Q$.
Until the mini black holes evaporate, an
energy $m_{\rm Planck}-H$ is confined to the internal degrees of freedom of the
three-tori. Of course, the presence of this state reflects the conceptual
difference between the $\Theta$ and $P$ sector on $Q$.

This identification, if realized, is a special one in that it may violate T
invariance. That is, the topological identifications in time need not
constitute diffeomorphisms. In fact, the holonomy of space-time is
reduced from the standard $O(4)=SU(2)\times SU(2)$ form to a single $SU(2)$ if
this degree of freedom is excited. This change in holonomy sounds rather
disastrous, but one should realize that even though this topological freedom
must be accommodated by Nature based on the thought experiment, its realization
through some field phenomenon is confined to quite narrow bounds.

To access this internal degree of freedom one requires a
particle/anti-particle system where both particles interact through a common
decay route or a neutral multiplet. Neutrality is a topological requirement due
to the $U(1)$ on the $L(T^3)$ junctions. The energy difference
between the rest masses of their superposition states should be equal to an
integer multiple of $F$
(resonance) in order to excite the vibration. It is well known that the
difference in weak self-energy determined by the superposition states
$$K_S\leftrightarrow 2\pi\leftrightarrow K_S\eqno(24a)$$
and
$$K_L\leftrightarrow 3\pi\leftrightarrow K_L,\eqno(24b)$$
for the decay of the $K^0$ and $\bar{K}^0$ mesons, is extremely small and
equal to $f=3.520\times 10^{-6}$ eV. Indeed, $f\approx F$ to less than one
percent. It is this coincidence which can allow the (indirect) CP violating K
meson decay processes to occur through $K_L\rightarrow K_S\rightarrow 2\pi$.
This conclusion demands the validity of CPT invariance. The possibility of a
CP violating process results from a topological identification in time, which
violates T invariance if the particles involved are not part of a multiplet,
i.e.\ topologically distinguishable on $L(T^3)$. Now if CPT is preserved in
kaon decay (as is indicated experimentally), then T violation in fact requires
the CP violating decay of the K meson. It is not clear whether CPT should be
preserved under all circumstances.

The level of violation (experimentally $0.227$\% for two-pion decay) is not
predicted by this argument, but is related to the excess in $F$ over $f$.
The relative
difference between $F$ and $f$ is $E=0.738$\%. For the intrinsic dispersion
$d=7^{-3}$ of the $T^3$ lattice vibration, and a sum of exponential decay rates
$p=\Sigma_{m=1}^{\infty}{\rm exp}^{-E/d}/7^{2m}$ per pair of three-tori with
seven degrees of freedom each and for all indirect routes, one finds
$R_{\rm CP}=p=0.162$\% (and an amplitude $p^2$ for direct routes through loop
identification of $K_L$ and $K_S$). This is in reasonable agreement with
experiment, given the use of a simple exponent for the barrier penetration
rather than $q^\mu q_\mu$ at the time of GUT symmetry breaking to determine the
probability for a certain fluctuation to occur {\it anywhere}, which reflects
the global deviations from the vacuum. Also, an additional coupling to $U(1)$
will lead to CP asymmetries of $1/7=14.3$\%.

For the neutrino multiplet similar results as above hold for a direct route,
i.e., with an effective mixing of $1/7^2=2.04$\%. It should be noted as well
that the uncertainty principle implies a characteristic time scale
$t_{\rm c}=7^3 (2\pi\hbar /F)$ during which different temporal identifications
are sustained in the neutrino sector, leading to an echo of some decay process
i$\rightarrow$f. The relative amplitude of this anomalous signal is determined
by the fraction of the energy available to the neutrino(s) multiplied by the
probability for decay of the initial state within $t_{\rm c}\approx 0.4\mu s$.
The echo occurs with a delay of $7t_{\rm c}$ relative to the transition
i$\rightarrow$f because of the seven-fold multiplicity of $T^3$. In this the
uncertainty principle causes the echo to spread out in time over more than
$t_{\rm c}$ as it is measured with higher and higher accuracy.

\section{A Spectral Observation of Planckian Topology through the Vacuum Energy}

\subsection{Spectral Lines}

Above it is shown that $L(T^3)$ possesses energy levels corresponding to
integer multiples $m$ of $E_0=2\pi\hbar\nu_0=F$
$$\nu_m =857.3588m\quad {\rm MHz},\eqno(25)$$
or $34.96698/m$ cm, where the numerical accuracy is limited by the Planck mass.
Energy considerations then
imply that photons of the appropriate wavelength can also interact with these
levels on any $U(1)$ lattice junction since they are massless.

A photon of the required energy will then
have a partner whose frequency $\nu_m'$ obeys $\nu_m'=\nu_0/m$. This is a
fixed point pairing under $E_0$ of $k^\mu k_\mu =0$ for the four vector
$(2\pi\nu_0,k_i)$. That is, for every
$\nu_m$ one should find a ghost image at frequency $\nu_m/m^2$ with a
wave vector differing only in scale. The duality
implies that destruction of either partner leads to destruction of the pair.
Furthermore, the pairing is not induced by photons with energies below $E_0$
since the suggested effect is quantized on the natural numbers.
Clearly, the distinguishing character of this signal resides in the fact that
its strength is always equal to that of the ``source'' spectrum at the high
frequency. At the frequency $\nu_0$ one can observe no effect, since detection
of the one photon implies removal of the partner. Furthermore, the high
(and low) frequency spectral features are also special in that their line
width is always equal to
$7^{-3}\nu_0=2.4996$ MHz (and $2.4996/m^2$ MHz). The intrinsic
line profile shape follows again from the mass-energy / particle rest mass
distribution of the entire Universe relative to the vacuum, derived from
$q^\mu q_\mu$ at the time of GUT symmetry breaking.
Finally, an elementary investigation of the change in free-free absorption
optical depth (at frequencies dual to the sun's 6000 K black body radiation
peak) associated with the sun's sunspot cycle yields correlated variations in
the sun's bolometric luminosity of the order of one part in a thousand because
long wavelength dual partners are absorbed.

\subsection{The Cosmological Constant: Vacuum Energy}

The zero-point $F$, as well as being the characteristic energy of
the lattice vibration, is a vacuum energy which behaves like a cosmological
constant. Its contribution is $\Lambda_F\sim 10^{-58}$ GeV$^4$.
To summarize the results on $\Lambda$, the TD of space-time
leads to quantities which behave like a cosmological constant through black
hole singularities and $L(T^3)$. Indeed, in the context of TD there are two
manifestations of $\Lambda$, but one underlying topological structure.

\section{Conclusions and Discussion}

A thought experiment has been proposed which leads to the notion of
three-tori and handles as fundamental objects on the Planck scale,
embodying the interplay between general relativity and quantum mechanics.
Together these prime manifolds form a fundamental topological manifold which
yields Lorentz invariance.
The general equation of motion has been derived for a possible QGUT on this
manifold $Q=2T^3\oplus 3S^1\times S^2$, which naturally leads
to a Higgs field, inflation, and the amplitude of the primordial
density fluctuations. The manifold
$Q$ contains the necessary symmetry groups to reproduce the standard model.
It possesses intrinsic energy scales which
determine the (position dependent) masses of all particles and the
values of the coupling constants, all of which are in good agreement with
current experimental limits. The spatial dependence of particle rest masses
on cosmological scales, with a $7^{-3}$ amplitude, leads to a gravitational
bias for mass-energy overdensities in the early Universe.
Specific, and easily falsifiable, predictions have been made
for, e.g.\ the mass of the Higgs boson and neutrinos,
and for a discrete spectral feature associated with the zero point
energy of the vacuum. These predictions make TD, which is clearly rather
ambitious, testable in a manner such that a single discrepancy rules it out,
i.e.\ TD contains no adjustable dynamical parameters.

The only three numbers which are unconstrained in TD are the speed of light,
the Planck length, and the Planck mass. This is not surprising since they
fix the scale of physical phenomena, but are not relevant for the nature of the
underlying physical system. In fact, the role played by topology in the theory
presented here suggests that Nature does not even distinguish physical reality
on this level, even though our perception of phenomena depends strongly on the
numerical values these three parameters attain.

Most importantly as far as the underlying philosophy of TD is concerned, it
was found that topological identifications play a crucial role in the way
Nature supports physical processes. Indeed, the essence of TD is that Nature
distinguishes all phenomena through geometric and topological information,
where the latter is not necessarily localized, as a reflection of the
Heisenberg uncertainty principle.

This raises the question of where to go next. If $Q$ and $L(T^3)$ turn out to
be fundamental, a fact which is amenable to observations and experiment, then
one can argue that TD provides a formulation of Nature which relates all
physical phenomena to one underlying structure, and as such is a theory of
everything. Fundamentally more important is what $Q$ and $L(T^3)$ imply
{\it beyond} such a unification. That is, the very nature of $L(T^3)$ suggests,
as has been suspected by many, that quantum mechanics is even more mysterious
than its original conception almost a century ago would suggest. In fact, since
the topological identifications on $L(T^3)$ are dynamical, and are already
required in an essential manner for known physical processes, one should
determine what other consequences they might have.

\acknowledgments

The author is indebted to E.E.~van 't Hof, G.~van Naeltwijck van Diosne,
J.A.A.~Vogels,
W.G.~Berendse, and M.C.~Spaans sr.\ for valuable assistance.
This work was supported by NASA through Hubble Fellowship grant HF-01101.01-97A
awarded by the Space Telescope Science Institute, which is operated by the
Association of Universities for Research in Astronomy, Inc., for NASA under
contract NAS 5-26555.\bigskip\bigskip

\centerline{\bf APPENDIX: THE AXIOM OF CHOICE AND THE PREDICTION OF GRAVITY}\bigskip\medskip

First a topological exposition is given of the structure which underlies $Q$
and $L(T^3)$. Consider $R^\infty$ and three-dimensional topological subspaces
$R^3$ therein under two different partitions: $S^1$ ($L$) and $S^2$ ($M$).
These partitions represent the axiom of choice: Any point can be {\it selected}
as the intersection of two loops, and any two points can be {\it distinguished}
if two spheres surrounding them are disjunct. Three space dimensions are
required to allow linked and unlinked loops, which yields the composite
manifolds $T^3=S^1\times S^1\times S^1$ and $H=S^1\times S^2$, and their
connected sums. Note that the presence of the $S^2$ sphere introduces a notion
of holography. Furthermore, introduce a measure of location, embodied by
$q^\mu$, on $[R^\infty/R^3]_{LM}$, with nuclearity provided by $T^3$ and $H$.
The partitions $L$ and $M$ are conjugate in that under homotopy, $S^1$ and
$S^2$ are related through two-point identifications.
One can now define unit elements, representing the possibility of choice,
through the dual action $S(L)\cdot S^{-1}(M)=1$, which expresses, in a unified
manner, the equivalence of inertial and gravitational mass and the Heisenberg
uncertainty principle. Hence, the number of paths through both partitions of a
unit element are equal, yielding $Q$.
The partitions $L$ and $M$ both reflect continuous (space-time) and discrete
(particle) degrees of freedom under $Q$.

The mathematical description above is the answer to the following question:
Is it possible, for a given set of objects, to choose an object from that set,
and to compare it to another object chosen from that set, through a choice
process that is defined solely in terms of the objects in the set itself?
If such a construction can be found then it is self-consistent in that it, and
the subsequent theory, is independent of any background, like a fixed
space-time or an external observer.

The first paragraph of this appendix describes this construction ($Q$) in a
recursive manner, as follows.
Imagine a collection of objects and possible connections, in a graph sense,
between them. Selection is a nontrivial link between two ring graphs in three
dimensions, under continuous deformation, whereas distinction follows from the
possible construction of two disjunct polyhedrons around two selected objects.
It then follows immediately that only the topology of $Q$ possesses the
selection and distinction properties, and is connected.
In fact, an elementary construction with triangles and tetrahedons directly
reproduces the number of degrees of freedom of $Q$ derived in the main text.
To be self-contained, each object in the original set should be viewed as a
collection of objects, with the same $Q$ topology, and this procedure is to be
repeated ad infinitum, i.e., to the continuum limit, to allow for invariance
under all the possible three-dimensional homeomorphisms that are required to
define the choice process. That is, this way the set of all possible
three-dimensional homeomorphisms on $Q$ does not require any choice of ordering
itself, and any object or collection of objects is amenable to choice, i.e.,
can be selected and distinguished. In this, any given object can be interpreted
as a point or a collection of points, and $Q$ favors no specific choice for the
numerical values of $\ell_{\rm Planck}$ and $m_{\rm Planck}$ at any given
point. From the above it appears that cause and effect should be viewed as the
equivalent sides of one coin.

Finally, the interchange symmetry of
the two identical three-tori in $Q$ yields the scalar $q^\mu q_\mu$ and the
symmetric tensor $C^{\mu\nu}=q^\mu q^\nu =C^{\nu\mu}$, which describes a spin 2
field, with four degrees of freedom. This spin 2 field constitutes the force
of gravity that couples to {\it any} mass-energy on $T^3$ because the $Z_2$
symmetry acts on all fields and their couplings on a three-torus.

\end{document}